\documentclass[aps,pre,
showpacs,showkeys,floatfix,superscriptaddress,twocolumn]{revtex4-2}
\usepackage[utf8]{inputenc}
\usepackage{graphicx}
\usepackage[caption=false]{subfig}
\usepackage[export]{adjustbox}
\usepackage[percent]{overpic}
\usepackage{natbib}
\usepackage[dvipsnames]{xcolor}

\begin{document}

\title{Critical behavior of cascading failures in overloaded networks}

\author{Ignacio A. Perez}
\email{Correspondence: ignacioperez@mdp.edu.ar}
\affiliation{Instituto de Investigaciones F\'isicas de Mar del Plata (IFIMAR)-Departamento de F\'isica, FCEyN, Universidad Nacional de Mar del Plata-CONICET, De\'an Funes 3350, (7600) Mar del Plata, Argentina}
\author{Dana Ben Porath}
\email{Formerly Dana Vaknin}
\affiliation{Faculty of Engineering and the Institute of Nanotechnology and Advanced Materials, Bar Ilan University, Ramat Gan 52900, Israel}
\author{Cristian E. La Rocca}
\affiliation{Instituto de Investigaciones F\'isicas de Mar del Plata (IFIMAR)-Departamento de F\'isica, FCEyN, Universidad Nacional de Mar del Plata-CONICET, De\'an Funes 3350, (7600) Mar del Plata, Argentina}
\author{Lidia A. Braunstein}
\affiliation{Instituto de Investigaciones F\'isicas de Mar del Plata (IFIMAR)-Departamento de F\'isica, FCEyN, Universidad Nacional de Mar del Plata-CONICET, De\'an Funes 3350, (7600) Mar del Plata, Argentina}
\affiliation{Physics Department, Boston University, 590 Commonwealth Ave., Boston, Massachussetts 02215, USA}
\author{Shlomo Havlin}
\affiliation{Department of Physics, Bar-Ilan University, Ramat Gan 52900, Israel}
\affiliation{Physics Department, Boston University, 590 Commonwealth Ave., Boston, Massachussetts 02215, USA}

\begin{abstract}
    \noindent
    
    While network abrupt breakdowns due to overloads and cascading
    failures have been studied extensively, the critical exponents and
    the universality class of such phase transitions have not been
    discussed. Here, we study breakdowns triggered by failures of links
    and overloads in networks with a spatial characteristic link length
    $\zeta$. Our results indicate that this abrupt transition has
    features and critical exponents similar to those of interdependent
    networks, suggesting that both systems are in the same universality
    class. For weakly embedded systems (i.e., $\zeta$ of the order of the
    system size $L$) we observe a mixed-order transition, where the order
    parameter collapses following a long critical plateau. On the other hand,
    strongly embedded systems (i.e., $\zeta \ll L$) exhibit a pure first-order
    transition, involving nucleation and the growth of damage. The
    system's critical behavior in both limits is similar to that observed
    in interdependent networks.
\end{abstract}

\maketitle

\section{Introduction}

Cascading failures and system collapse due to overloads have been modeled
and studied within a network framework \cite{mott-02,mott-04}. Even a small
failure (e.g., deliberate attacks, natural disasters, or random malfunctions)
may spread the overloads in relevant infrastructure such as power grids, 
transportation networks, and communication systems, producing a partial or 
total collapse. Thus, understanding the laws of cascading failures due to 
overloads (CFO) is crucial for ensuring the operation of infrastructure and
services that we rely on every day. Infrastructure is often embedded in two- 
or three-dimensional space \cite{watt-98,bart-11,gros-17,gros-22-b} and, far
from ideal systems such as lattices, many real-world networks present a 
characteristic link length $\zeta$ \cite{wax-88,daq-11,2012japan}. Several 
studies \cite{dan-16,gros-17,vak-17,perez-22,gote-22} model this property 
with a two-dimensional (2D) lattice where the sites are the nodes of the 
network and link lengths are chosen from an exponential distribution, 
$P(r)~\sim~exp(-r/\zeta)$ (the so-called $\zeta$ model), allowing dimension 
to change from two, for small $\zeta$ (short links), to infinite for large
$\zeta$ (i.e., of order of the system linear size $L$) \cite{daq-11}. Thus,
in the $\zeta$ model, the parameter $\zeta$ represents the strength of the 
spatial embedding.

A fundamental model for CFO is the one developed by Motter and Lai
(ML)~\cite{mott-02} that introduced the concept of load and overload for a
node. In this model, load is defined as the number of shortest paths that
pass through the node, and is considered a measure of relevance in the
transmission of some quantity (e.g., information or energy) throughout the
system. They also defined a threshold called capacity, proportional to the
initial load and representing the maximum load that a node can hold. Above
the capacity, the node becomes overloaded and fails. However, the shortest
path is not always the optimal path~\cite{havl-05}. A reasonable modification
of this model is defining weighted networks, where links have associated
weights that may indicate, for instance, the time (or cost) that it takes
to traverse a link. In this way, optimal paths, which represent the paths
with minimal travel time (or cost) between nodes, are considered to define
the loads.

Currently, the critical behavior and the universality class of CFO's phase
transition have not been systematically studied. Here, we study this phase
transition in both, spatial $\zeta$ model~\cite{dan-16} and in
Erd\H{o}s-R\'{e}nyi~(ER)~\cite{erdos-59,bunde-91,new-10} networks,
finding indications that it belongs to the same universality class as
percolation of interdependent 
networks~\cite{bul-10,gao-11,li-12,zho-14,dan-14,kia-21}. We observe that
for weakly or non-spatially-embedded systems, such as ER networks or the
$\zeta$ model for large $\zeta$ (i.e., $\zeta \sim L$), there exists a
mixed-order transition, similar to interdependent
ER~networks~\cite{bul-10,gao-11,zho-14,berez-15}. At this abrupt transition,
we find a long-term plateau in the order parameter characterized by critical
exponents. In contrast, for strongly embedded networks, (i.e., 
$\zeta~\ll~L$), we observe a pure first-order transition caused by
nucleation of a random damage, a behavior also exhibited by interdependent
lattices with finite-length dependencies or spatial multiplex 
networks~\cite{berez-15,gros-22}.

\section{Model}

Our system consists of a 2D-lattice of size $N =~L \times~L$ with
link lengths $r$ exponentially distributed, i.e., $P(r) \sim~exp(-r/\zeta)$
($\zeta$ model~\cite{dan-16}), and average degree $\langle k \rangle$
(self- or multiple connections are not allowed, and we assume periodic
boundary conditions). Regarding the CFO dynamics, we study the ML
model~\cite{mott-02} in weighted networks, with positive weights that
follow a Gaussian distribution. We define the load of node $i$,
$L_i(t) \equiv L_i^t$, as the number of optimal paths between all pairs
of nodes, excluding $i$, that pass through $i$ at time $t$. The maximum load
that a node can sustain at any time is given by its capacity,
$C_i = L^0_i(1 + \alpha)$, which is proportional to the initial load $L^0_i$.
The parameter $\alpha$ is the system's tolerance, and it represents the
resilience of nodes to failure.

\begin{figure}
    \captionsetup{skip=0.9cm}
    \subfloat{\begin{overpic}[width = 0.1\columnwidth, height = 0.25\columnwidth]{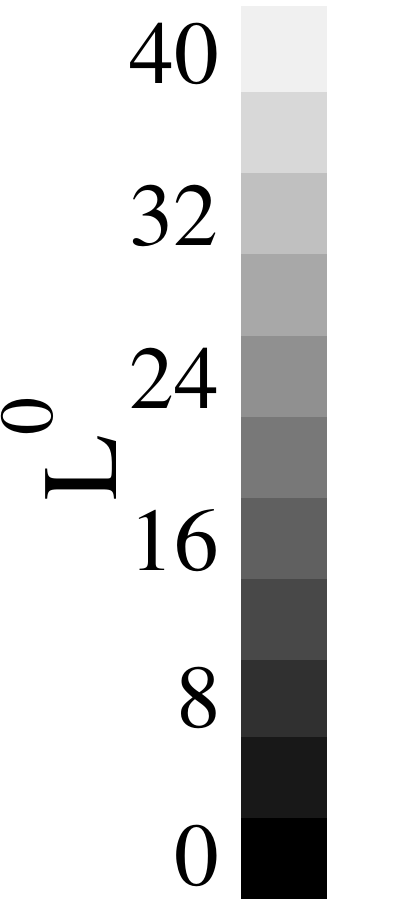} 
    \end{overpic}}
    \subfloat{\begin{overpic}[width = 0.25\columnwidth, height = 0.25\columnwidth]{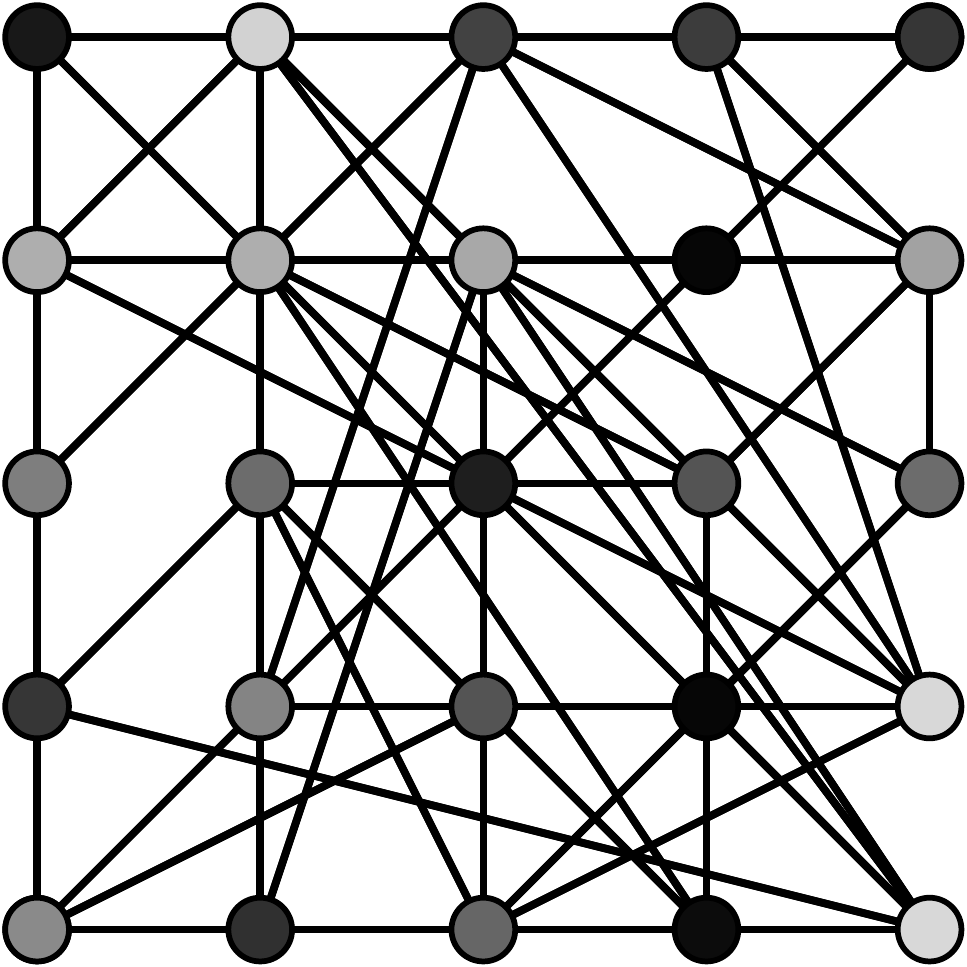} 
        \put(0,-15){{\bf(a)}}
        \put(35,-15){$t = 0$}
    \end{overpic}}
    \hspace{0.005\columnwidth}
    \subfloat{\begin{overpic}[width = 0.25\columnwidth, height = 0.25\columnwidth]{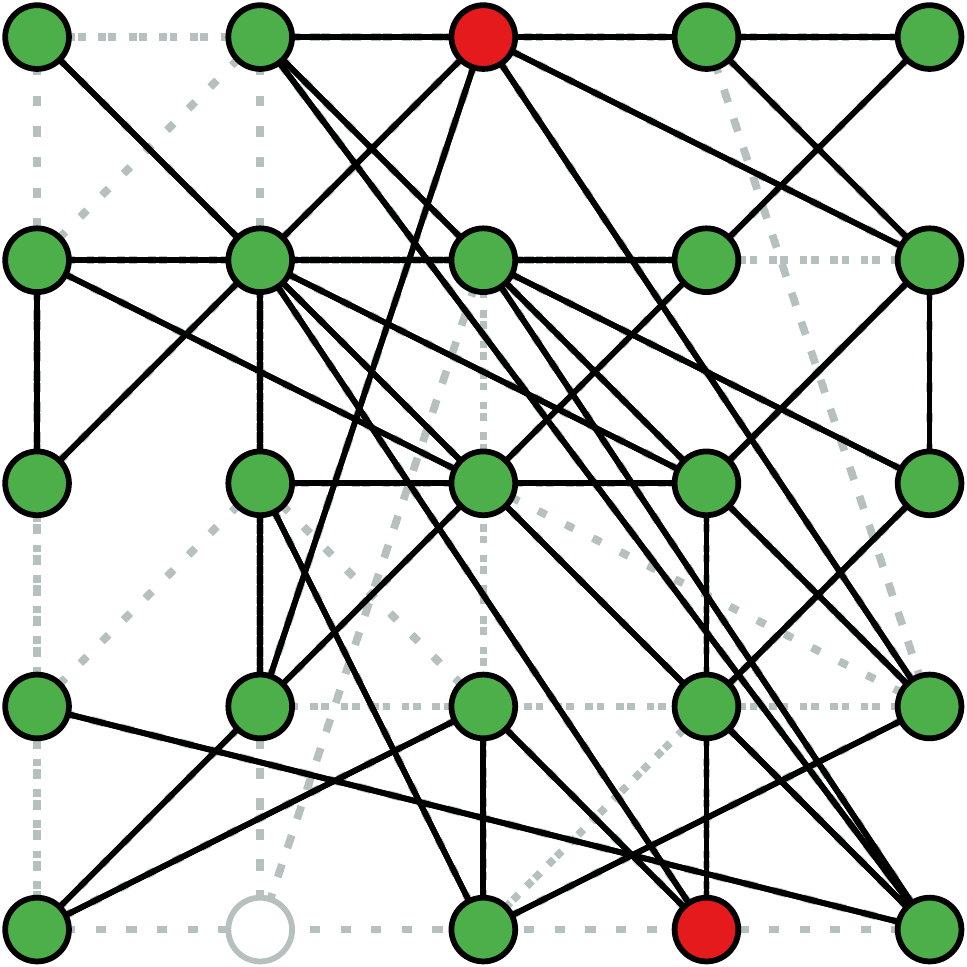} 
        \put(0,-15){{\bf(b)}}
        \put(35,-15){$t = 1$}
    \end{overpic}}
    \hspace{0.005\columnwidth}
    \subfloat{\begin{overpic}[width = 0.25\columnwidth, height = 0.25\columnwidth]{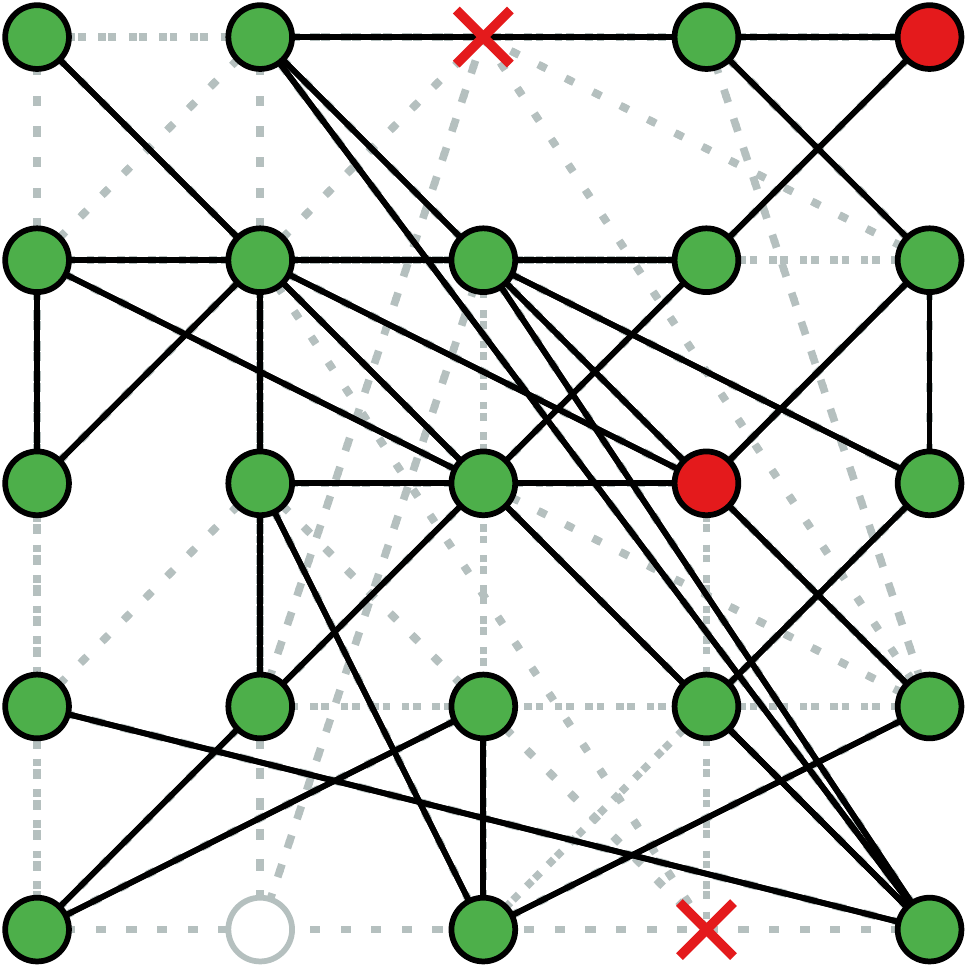} 
        \put(0,-15){{\bf(c)}}
        \put(35,-15){$t = 2$}
    \end{overpic}}
    \hspace{0.005\columnwidth}
    \subfloat{\begin{overpic}[width = 0.08\columnwidth, height = 0.25\columnwidth]{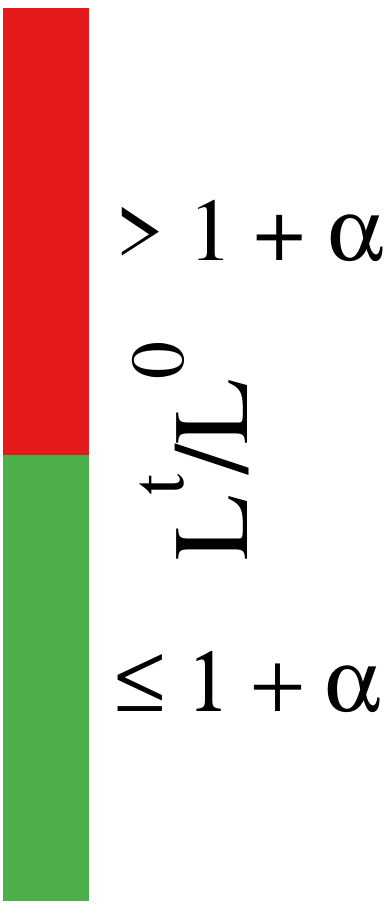} 
    \end{overpic}}
    \caption{Model demonstration. (a) A spatially embedded network
    with $L = 5$, $\langle k \rangle = 4$, and $\zeta = 5$. The load
    is represented by colors, increasing from dark to light black.
    (b) Randomly removed links ($1 -~p = 0.21$), represented by gray
    dashed lines, produce changes in the node's loads. For a green node $i$,
    $L_i^1/L_i^0 \leq 1 + \alpha$, i.e., the load does not exceed the node's
    capacity, while for a red node $j$, $L_j^1/L_j^0 > 1 + \alpha$, and
    the node becomes overloaded and fails. (c)~Failed nodes (red crosses)
    are removed altogether with their links, producing new overloads
    that continue the cascade process. The tolerance for this network is
    $\alpha = 2$.} \label{fig1}
\end{figure}

At $t = 1$, we randomly remove a fraction $1 - p$ of links, $p~\in~[0, 1]$.
This produces changes of the optimal paths throughout the network, affecting
the node's loads, which may generate successive failures due to nodes that become
overloaded, in a cascade manner (see Fig.~\ref{fig1}). After removing the links,
we advance one unit of time and compute the new loads. For $t > 1$, node $i$
fails if $L_i^t > C_i$, we remove failed nodes and their links, and advance one
unit of time. We repeat the process until there are no more failures in the
network.

The model presented above is not solvable analytically because of spatial
constraints, but it can be analyzed via extensive time-consuming numerical
simulations. To reduce the sensitivity of the results and produce smoother
and consistent curves for a single realization, randomness is somewhat
reduced. When performing percolation using a series of $1 - p$ values, we
proceed as follows: If $E_{p_1}$ is the set of links that have been
randomly removed for $1 - p_1$ then, for a larger value $1 - p_2$, we
remove the same set of links $E_{p_1}$ and additional random links until
we reach the value $1 - p_2$.

\section{Results}

\begin{figure}
    \subfloat{\begin{overpic}[width = 0.5\columnwidth, height = 0.4\columnwidth]{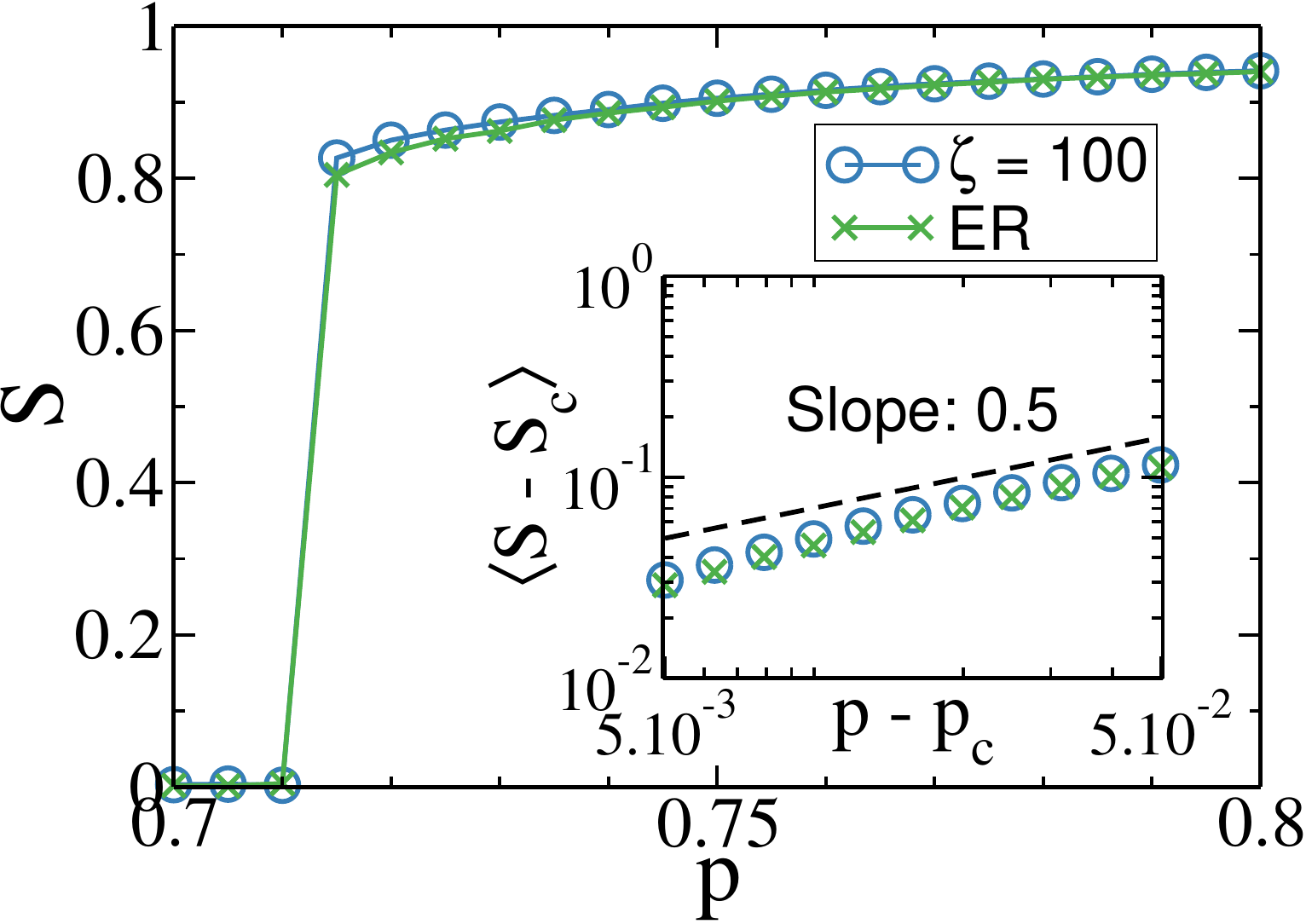} 
        \put(0,0){\bf{(a)}}
    \end{overpic}}
    \subfloat{\begin{overpic}[width = 0.5\columnwidth, height = 0.4\columnwidth]{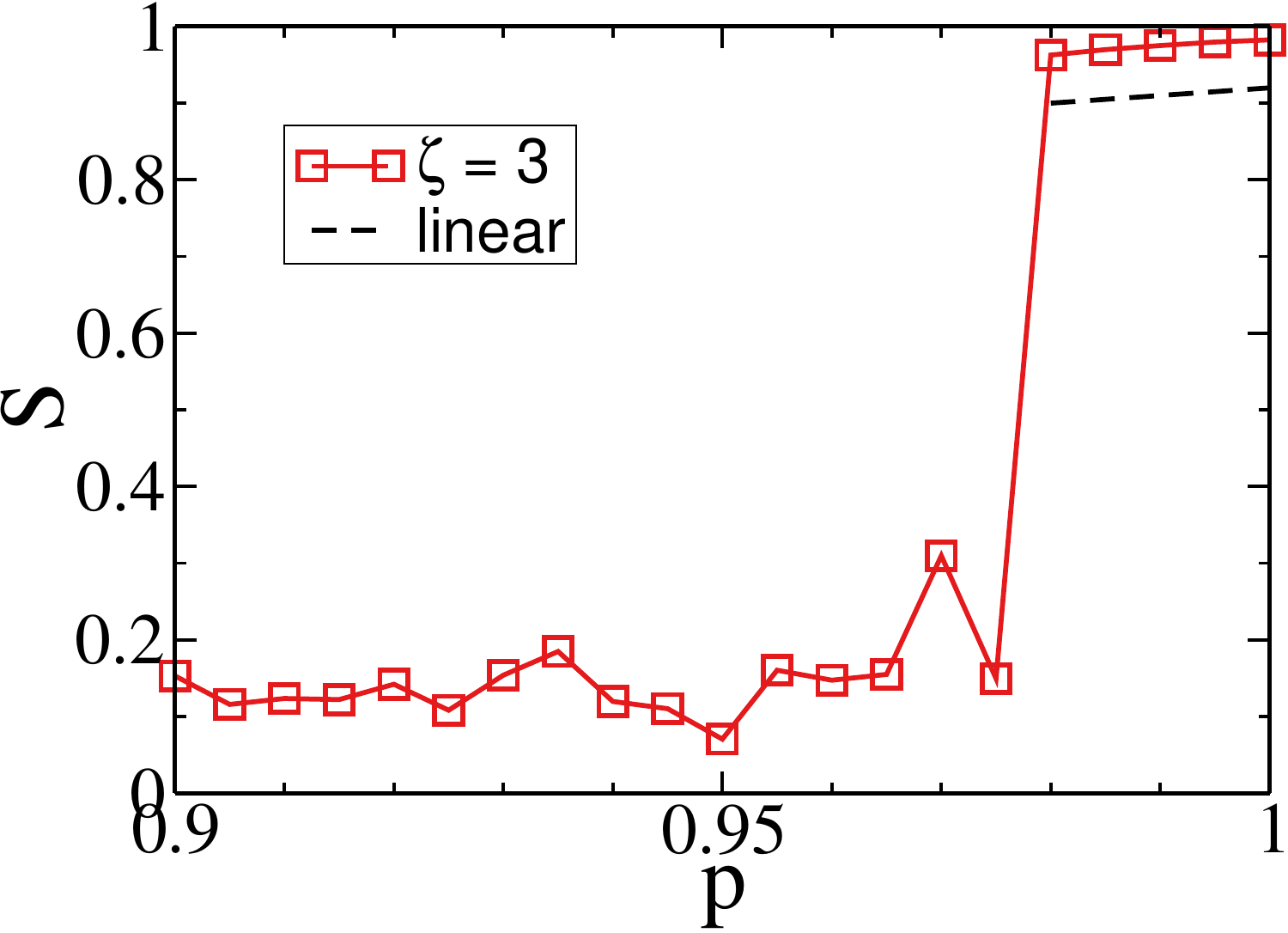} 
        \put(0,0){\bf{(b)}}
    \end{overpic}}
    \caption{Giant component of functional nodes, $S$, as a function of the
    fraction of non-removed links, $p$, for (a)~$\zeta = 100$ and
    (b)~$\zeta = 3$. In the inset of (a), we show a dashed line with slope
    $\beta = 1/2$, which characterizes the system's power-law behavior at
    criticality for large~$\zeta$. The slight deviation from a power-law (with
    exponent $1/2$) behavior observed when the system is close to criticality
    (left side of the inset) is due to finite-size effects, and should improve
    as we increase the linear size of the system, $L$. Results in the main plots
    correspond to individual realizations, while the inset plot is obtained by 
    averaging $S - S_c$ over 10 runs at each value of $p - p_c$. The remaining network parameters are $L = 350$ (for the ER network $N = 122500 = 350^2$), $\langle k \rangle = 4$, and $\alpha = 2$.} \label{fig2}
\end{figure}

We analyze the relative size of the giant component of functional nodes at
the end of the cascading process, $S(p) \equiv S$, for weak and strong 
spatial embedding, i.e., for large and small $\zeta$, respectively. This is
shown in Fig.~\ref{fig2}. In both cases, we find that the system undergoes an
abrupt transition at a critical value $p_c$, such that $S(p \geq~p_c) > 0$.
Nevertheless, we can distinguish two different behaviors at the vicinity of
these transitions.
For weak spatial embedding ($\zeta =~100$, Fig.~\ref{fig2} (a)) the
system approaches criticality, for $p > p_c$ and $S > 0$, with a clear
curvature that is absent for strong embedding ($\zeta = 3$,
Fig.~\ref{fig2}~(b)). We characterize the weakly embedded system through
a generalization of the critical exponent $\beta$ for abrupt
transitions~\cite{bocc-16,gros-22}, with respect to $S(p_c) > 0$: 
$S(p) -~S(p_c) \sim (p - p_c)^{\beta}$, $p \gtrsim p_c$. In the inset of
Fig.~\ref{fig2}~(a), we show that $\beta \cong 0.5$ for $\zeta = 100$,
coinciding with the usual mixed-order transition and with interdependent
random networks \cite{gros-22}. In contrast, for a strong spatial structure
($\zeta = 3$, Fig.~\ref{fig2} (b)), we do not observe a curvature with
a critical exponent, but just a linear decrease followed by an abrupt
collapse, suggesting a pure first order transition as in interdependent
spatial networks (see, e.g., Fig.~1 in Ref. \cite{dan-14}).

\begin{figure}
    \captionsetup{skip=0.8cm}
    \subfloat{\begin{overpic}[width = 0.5\columnwidth, height = 0.35\columnwidth]{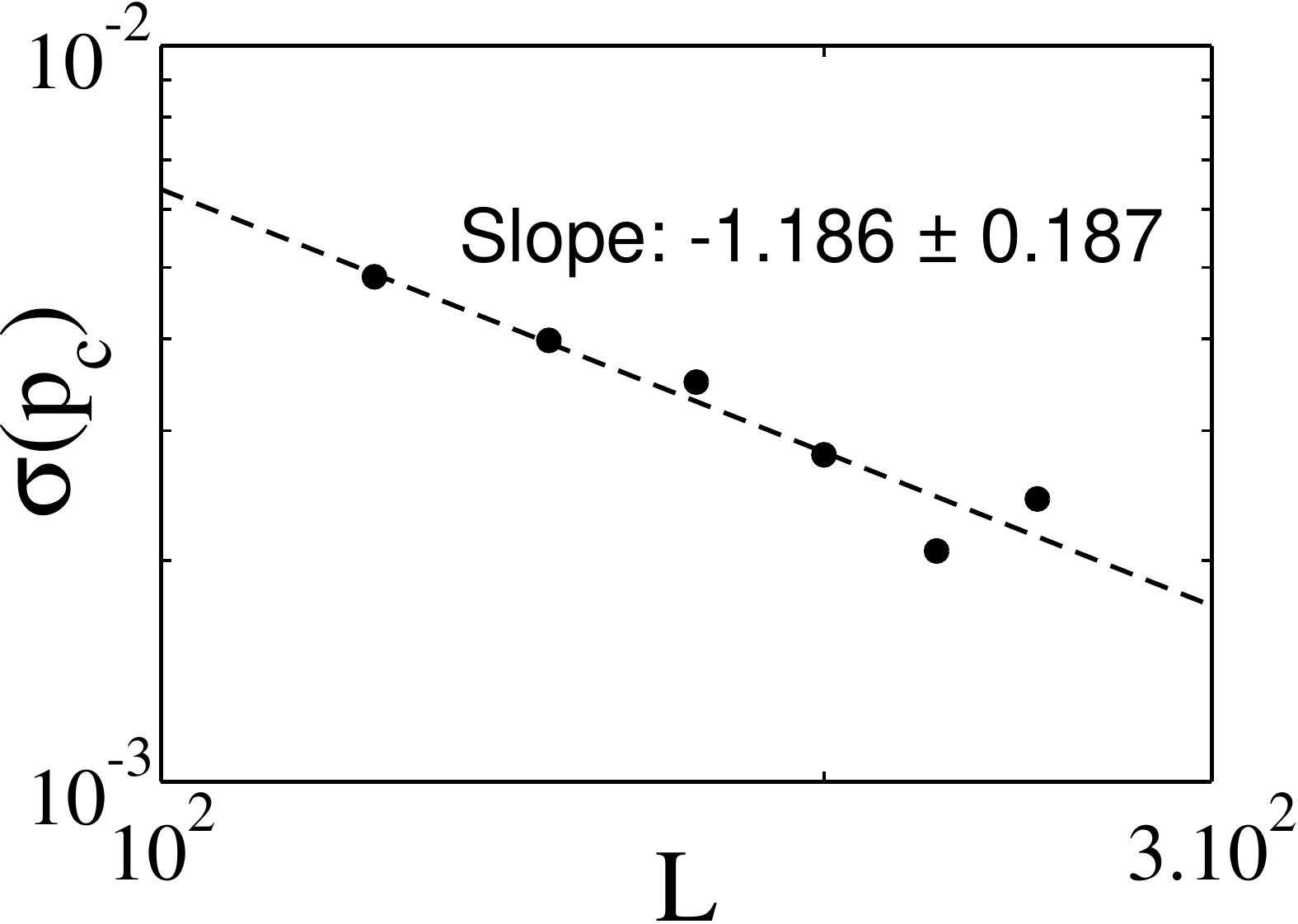} 
        \put(0,-5){\bf{(a)}}
    \end{overpic}}
    \subfloat{\begin{overpic}[width = 0.5\columnwidth, height = 0.35\columnwidth]{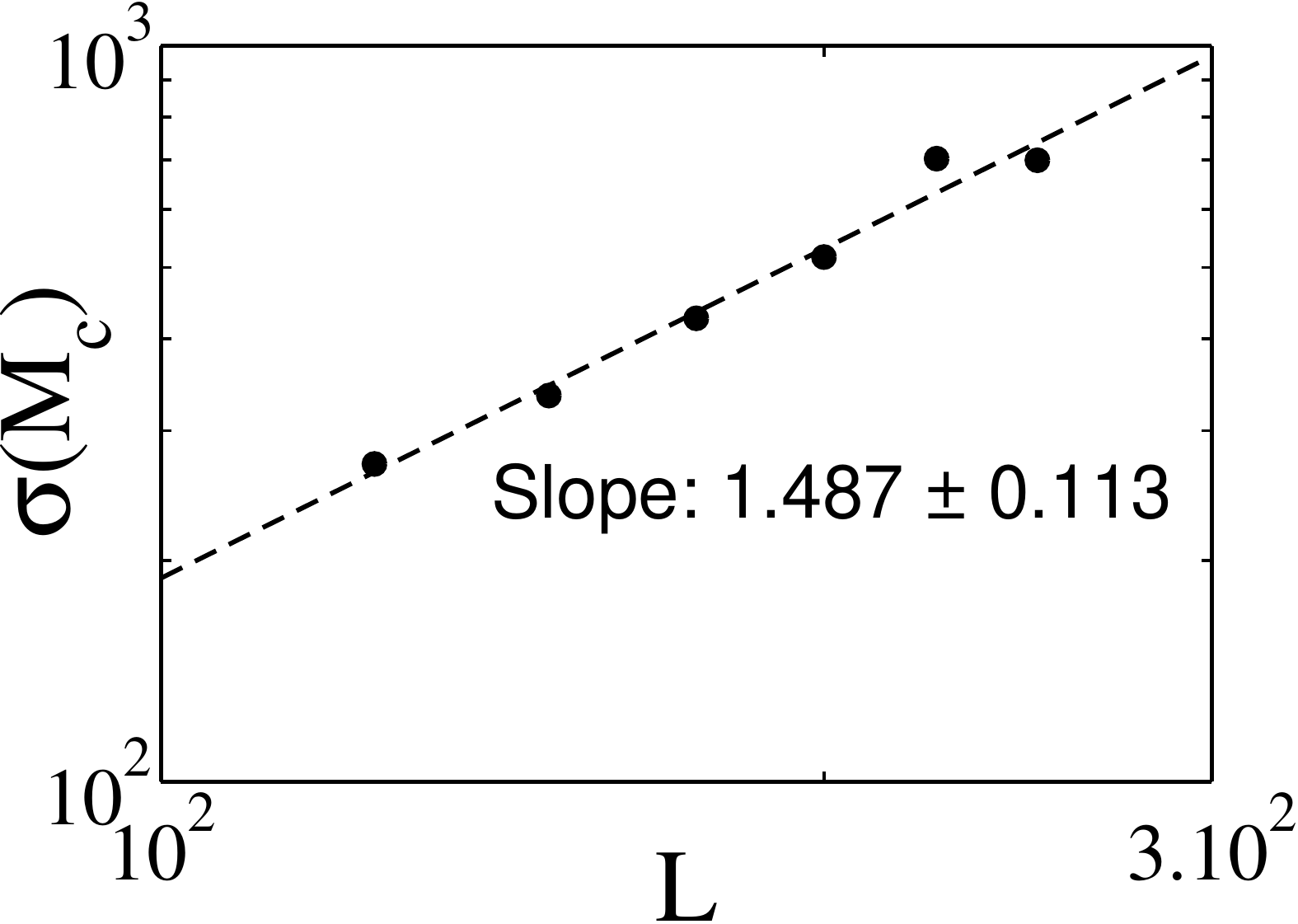} 
        \put(0,-5){\bf{(b)}}
    \end{overpic}}
    \caption{Fluctuations of (a) the critical threshold, $\sigma(p_c)$, and
    (b) the mass of the giant component at criticality, $\sigma(M_c)$. The
    dashed lines correspond to a power-law fit to the data, and we inform
    the slopes and their standard deviation. The scaling exponents obtained
    for mean-field networks with $\zeta = L$ are consistent with those evaluated for
    interdependent networks in two dimensions~\cite{gros-22-b}, 
    ~$\nu' \cong -1$ and $d'_f \cong 1.5$. Results for each value of $L$
    correspond to an average of $40$ independent realizations.} \label{fig3}
\end{figure}

The critical threshold $p_c$ and the mass of the giant component at $p_c$,
$M_c = N S_c$, may vary between realizations (see Fig.~1 of the Supplemental
Material \cite{sm}). We study their fluctuations, 
$\sigma(p_c) =~(\langle {p_c}^2 \rangle -~\langle p_c \rangle^2)^{1/2}$ and
$\sigma(M_c) =~(\langle {M_c}^2 \rangle - \langle M_c \rangle^2)^{1/2}$, on 
networks with long-range connectivity links ($\zeta = L$) and different
system sizes. Gross {\it et al.} ~\cite{gros-22-b} found for interdependent 
networks
with long-range dependencies that a finite-size scaling analysis yields
the relations $\sigma(p_c) \sim~L^{-1/\nu'}, \qquad \nu' = 2/d$, and
$\sigma(M_c) \sim~L^{d'_f}, \qquad d'_f = 3d/4$, where $d$ is the spatial
dimension. In Fig.~\ref{fig3}, we show that for the ML~overload 
model~\cite{mott-02} there exists a similar scaling with the linear
system's size $L$, and with the same exponents (i.e., for $d = 2$,
$\nu' = 1$ and $d'_f = 3/2$).

\begin{figure*}
	\captionsetup{skip=0.5cm}
    \begin{overpic}[width = 0.24\textwidth, height = 0.18\textwidth]{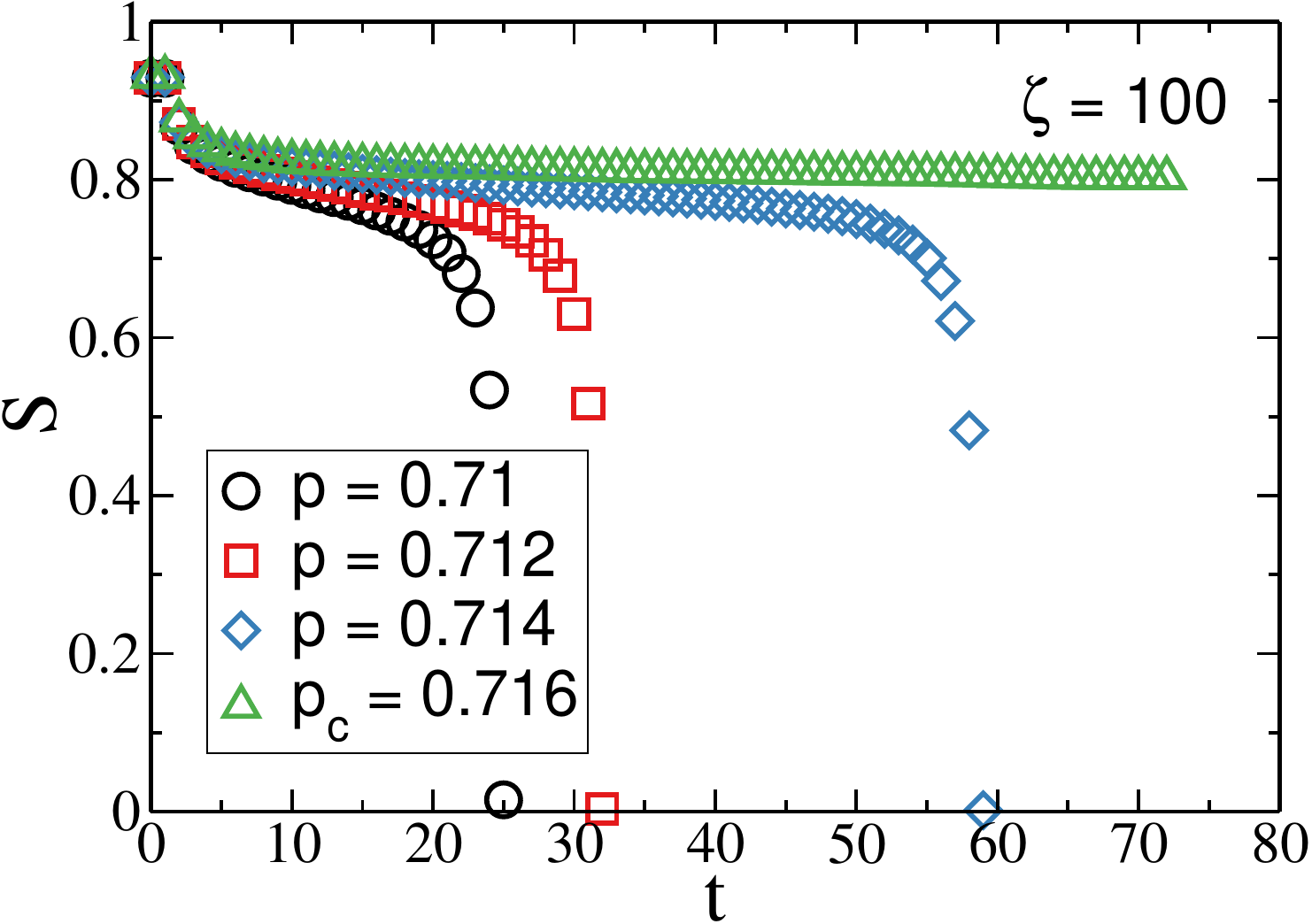} 
        \put(0,-5){\bf{(a)}}
    \end{overpic}
    \begin{overpic}[width = 0.24\textwidth, height = 0.18\textwidth]{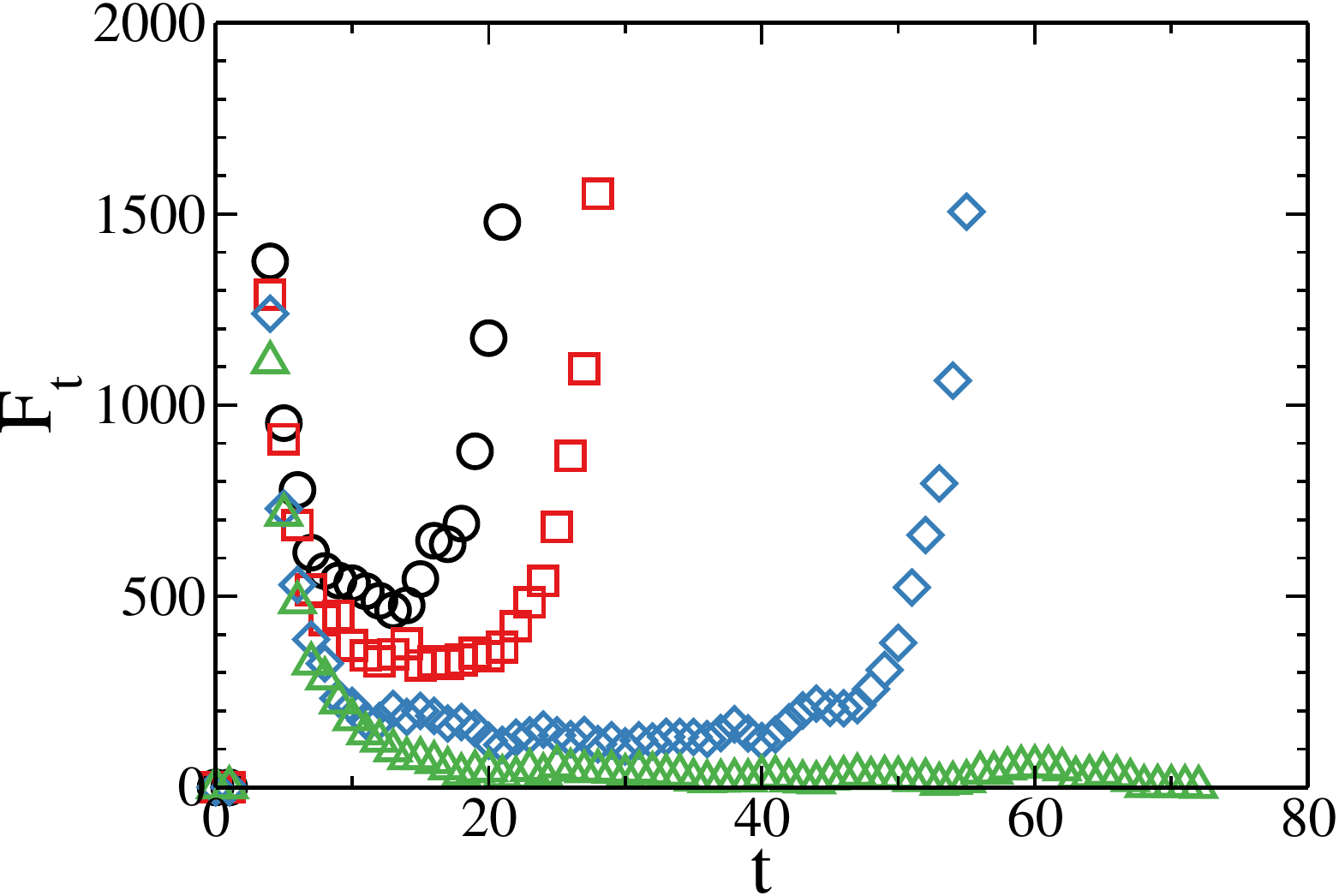} 
        \put(0,-5){\bf{(b)}}
    \end{overpic}
    \begin{overpic}[width = 0.24\textwidth, height = 0.18\textwidth]{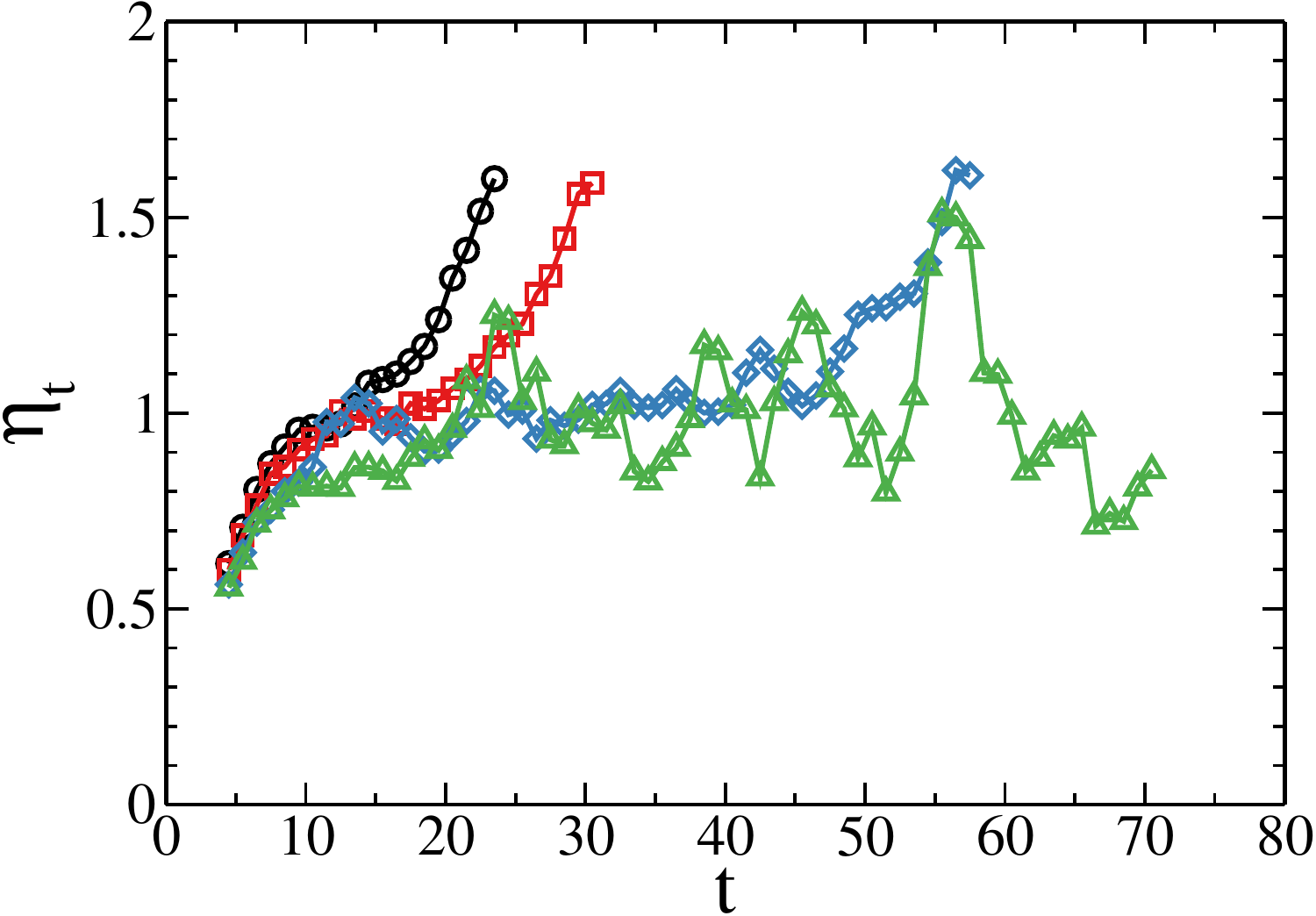} 
        \put(0,-5){\bf{(c)}}
    \end{overpic}
    \raisebox{0.15cm}{\begin{overpic}[width = 0.25\textwidth, height = 0.175\textwidth]{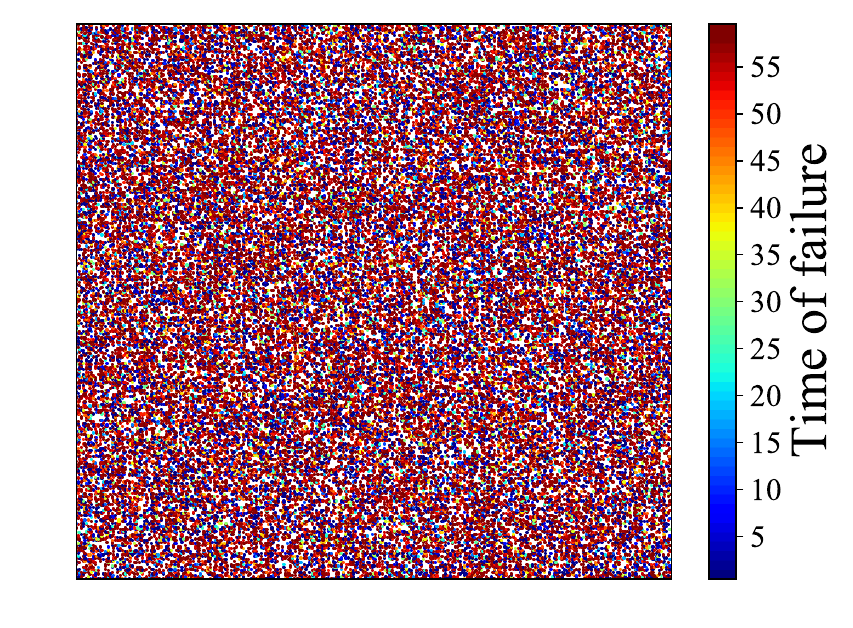} 
        \put(0,-8){\bf{(d)}}
    \end{overpic}} \\
    \vspace{0.3cm}
    \begin{overpic}[width = 0.24\textwidth, height = 0.18\textwidth]{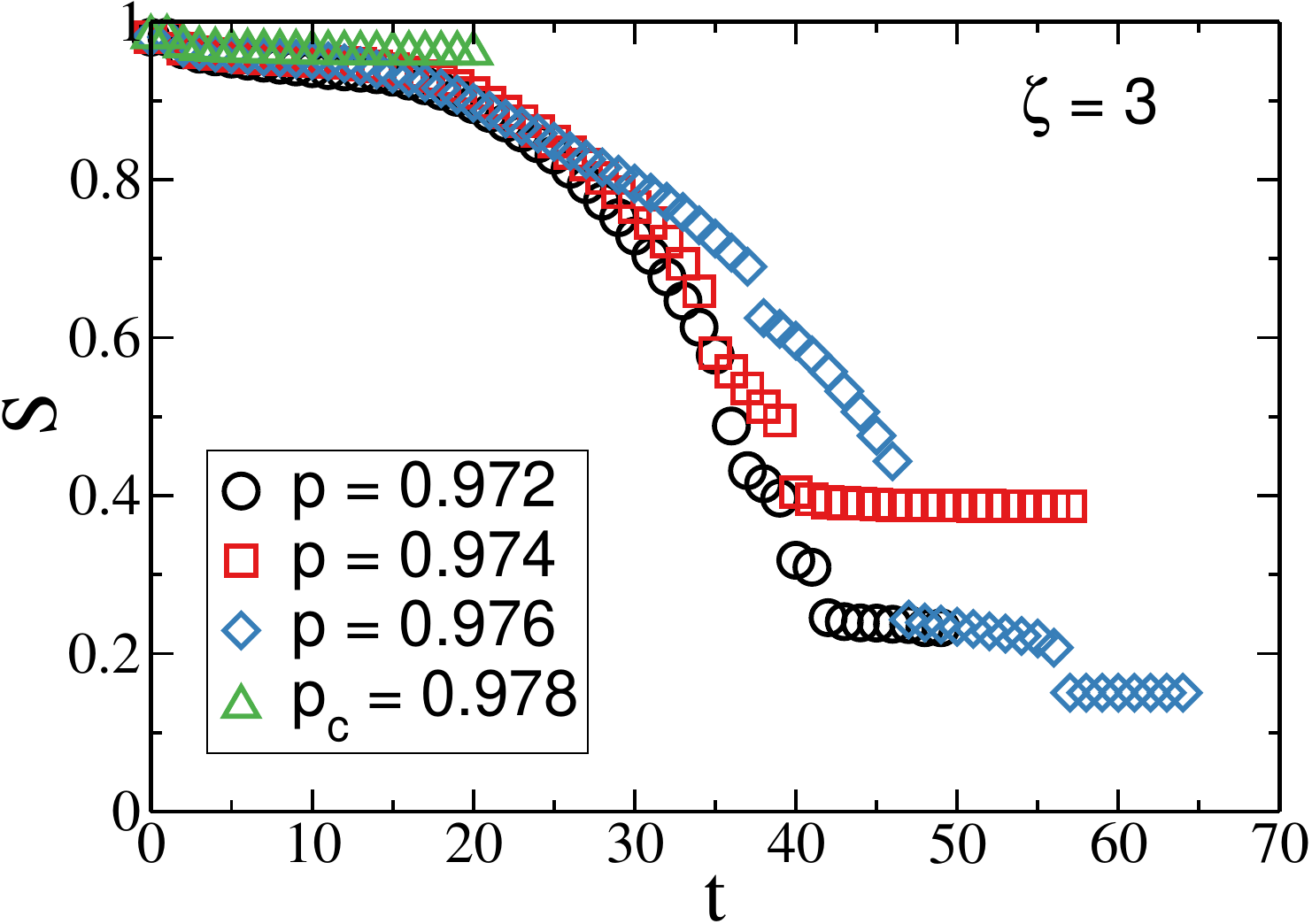} 
        \put(0,-5){\bf{(e)}}
    \end{overpic}
    \begin{overpic}[width = 0.24\textwidth, height = 0.18\textwidth]{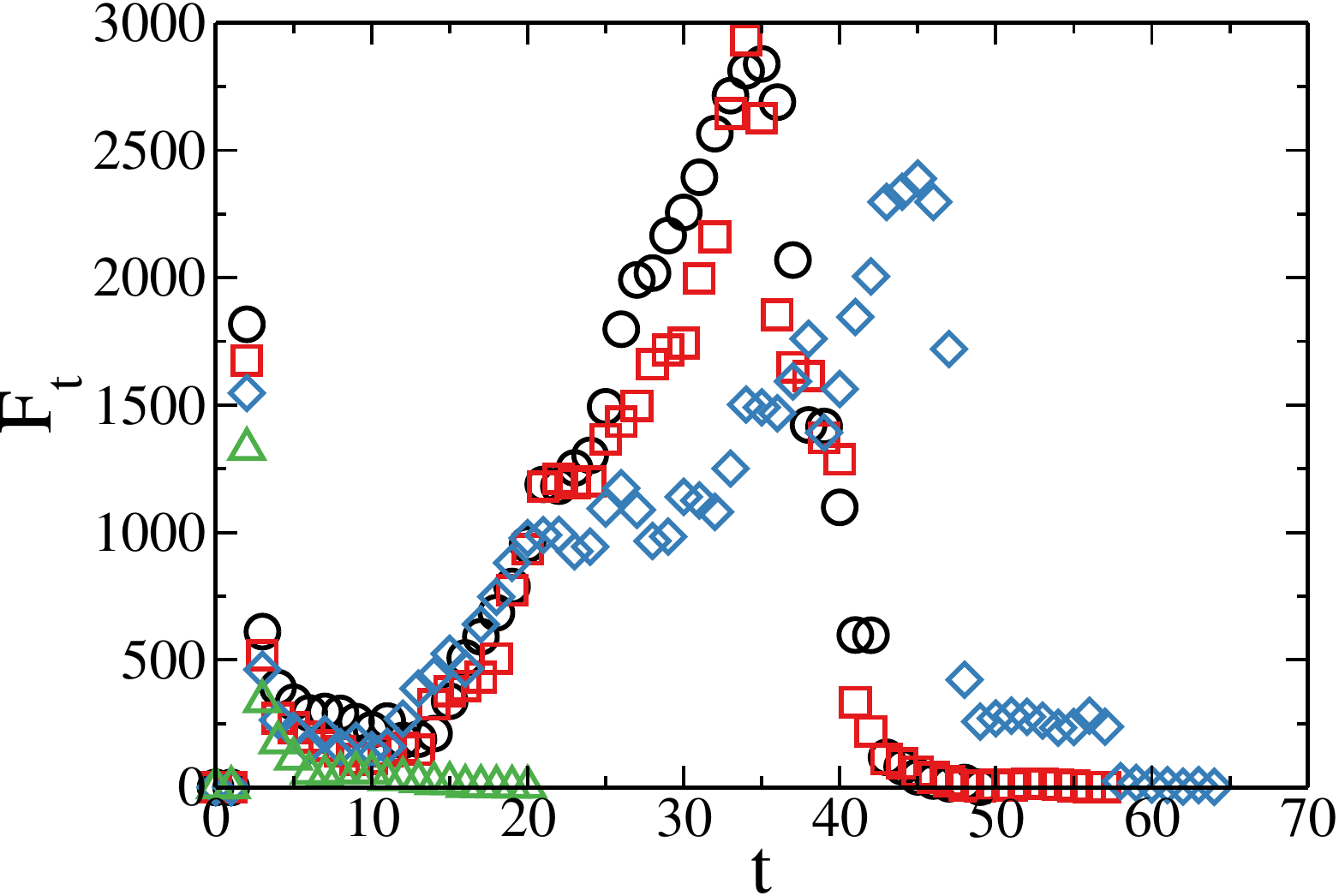} 
        \put(0,-5){\bf{(f)}}
    \end{overpic}
    \begin{overpic}[width = 0.24\textwidth, height = 0.18\textwidth]{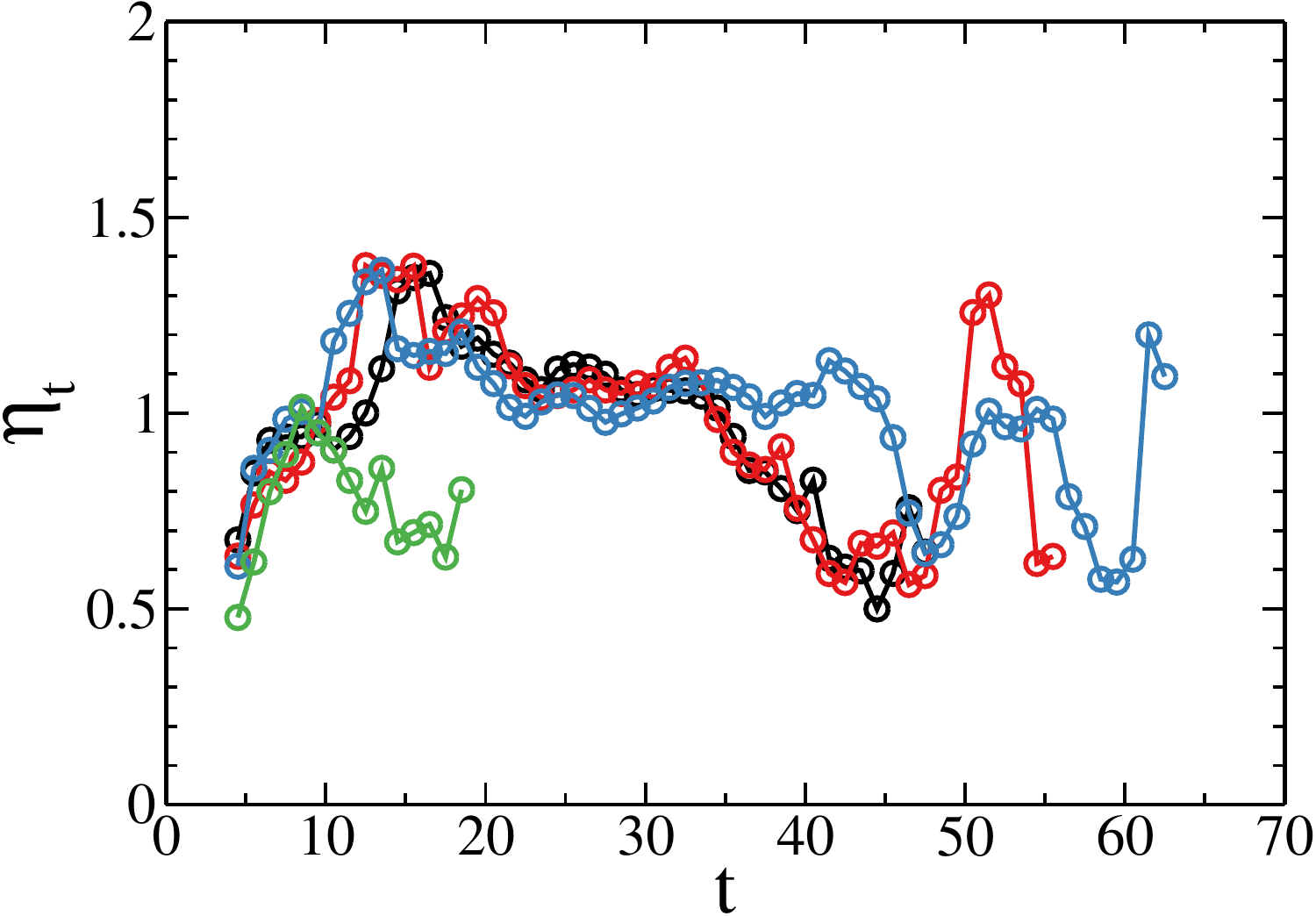} 
        \put(0,-5){\bf{(g)}}
    \end{overpic}
    \raisebox{0.15cm}{\begin{overpic}[width = 0.25\textwidth, height = 0.175\textwidth]{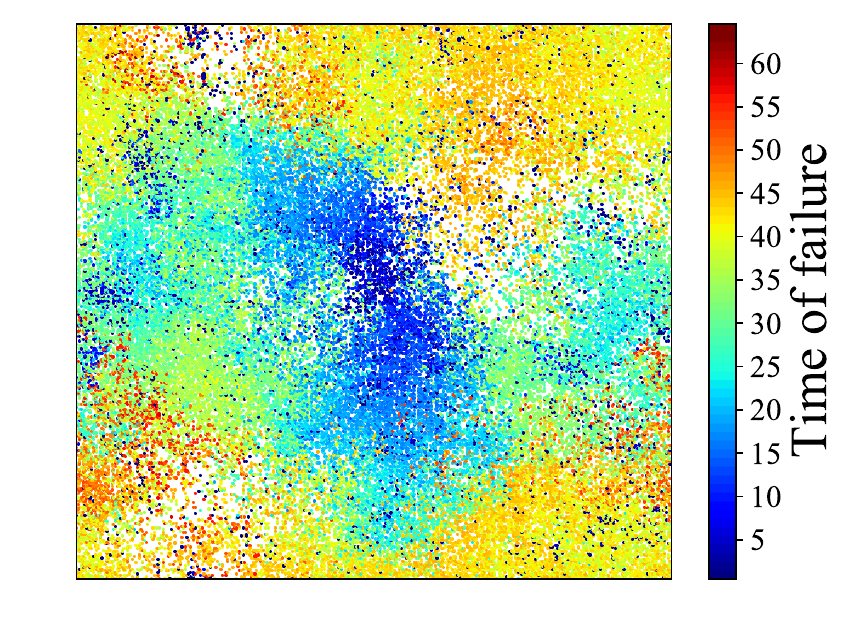} 
        \put(0,-8){\bf{(h)}}
    \end{overpic}}     
    \caption{Dynamic behavior of overload failures model near criticality,
    for networks with $\zeta = 100$ (top figures) and $\zeta = 3$ (bottom figures).
    From left to right: (a) and (e) Evolution of the giant component relative
    size $S$, with $t$ the number of iterations. (b)~and~(f)~Instant 
    failures, $F_t$. Note the microscopic values and the flatness increment
    in (b), in contrast to that in (f). (c)~and~(g) Moving average of the
    branching factor $\eta_t$. $\eta_t$ stays for a long time around $1$,
    indicating critical branching. (d)~and~(h) Spatiotemporal propagation of
    the failures. Colors represent the time of node failure close to criticality
    for (d)~$p = 0.714$ ($\zeta = 100$) and (h) $p = 0.976$ ($\zeta = 3$, the 
    colormap is centered to improve visualization). It is clearly seen in (h)
    that short-range connections in strongly embedded networks originate a 
    spatial-radial spreading of failures, in a process known as nucleation 
    (see also Ref. \cite{zhao-16}). This is in contrast to (d) where damage spreads
    to any location due to long-range links ($\zeta = 100$). Legends in (a)-(c)
    are the same, as well as in (e)-(g).} \label{fig4}
\end{figure*}

\begin{figure}
	\captionsetup{skip=0.8cm}
    \subfloat{\begin{overpic}[width = 0.5\columnwidth, height = 0.35\columnwidth]{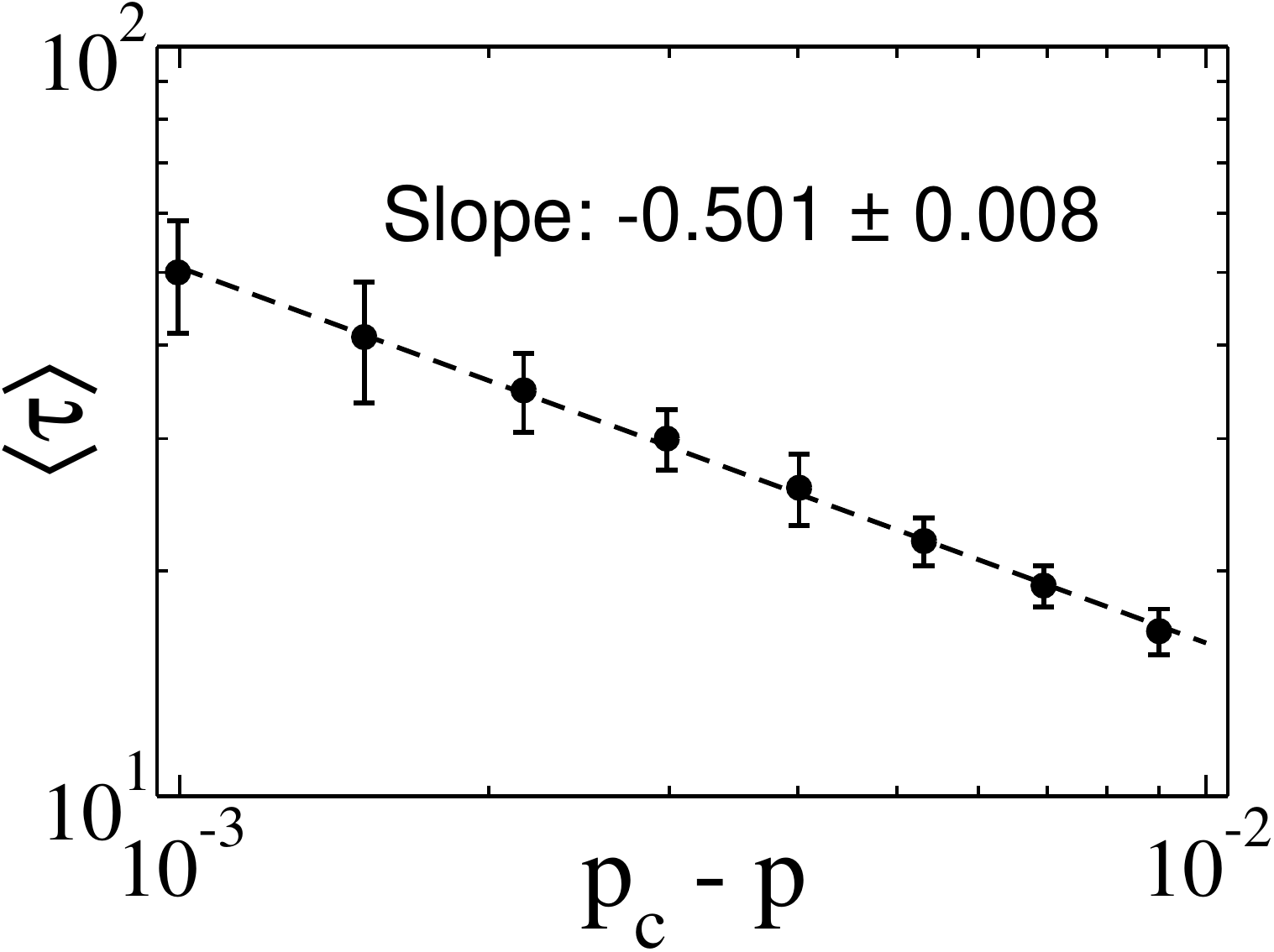} 
        \put(0,-5){\bf{(a)}}
    \end{overpic}}
    \subfloat{\begin{overpic}[width = 0.5\columnwidth, height = 0.35\columnwidth]{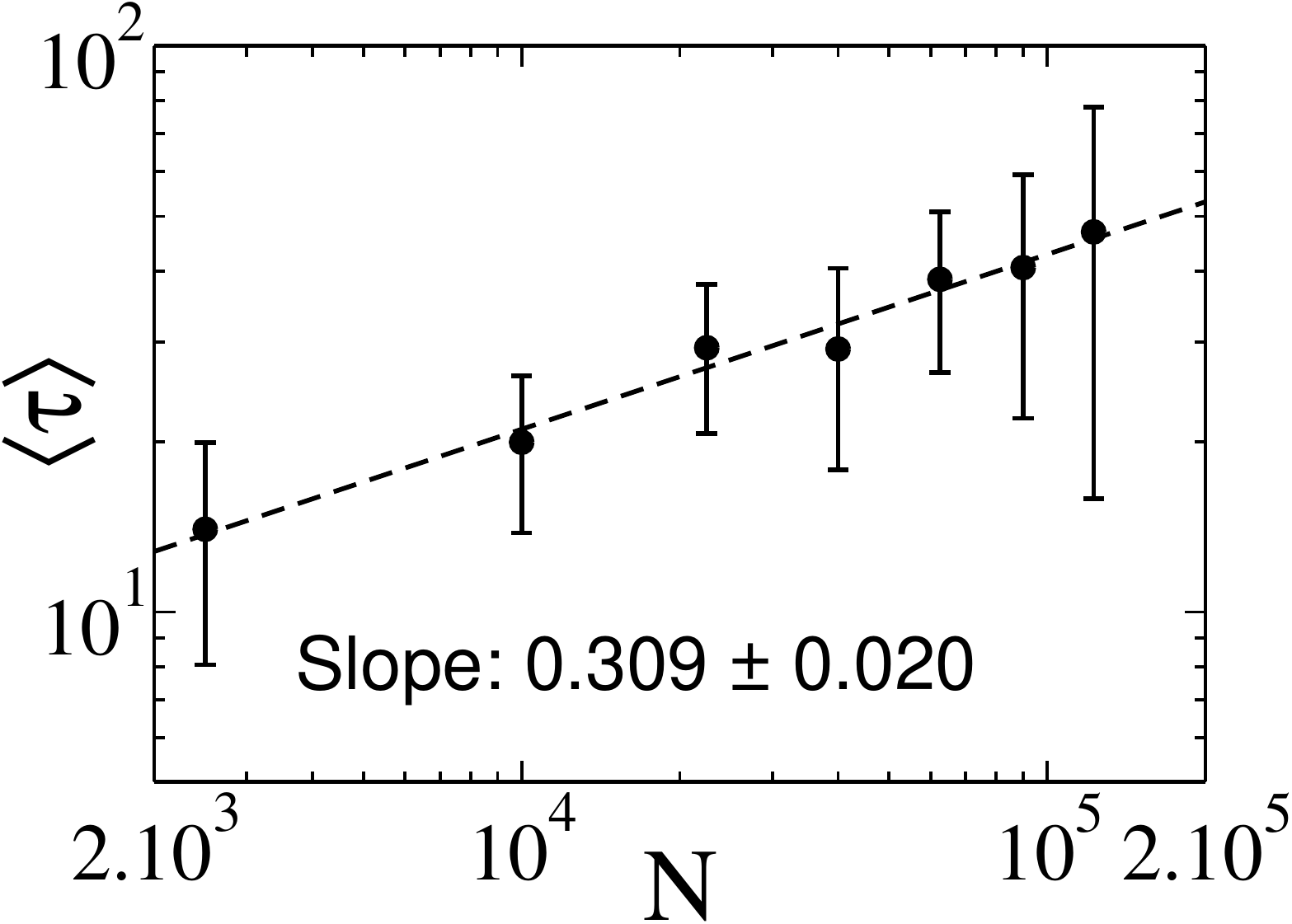} 
        \put(0,-5){\bf{(b)}}
    \end{overpic}}
    \caption{Scaling behavior of the average total time of the cascade,
    $\langle \tau \rangle$, for $\zeta = 100$. The dashed lines correspond to a
    power-law fit of the curves, and we inform the slopes and their standard
    deviation. (a)~$\langle \tau \rangle$ scales with $p$ near criticality, 
    $p_c - p$ ($p < p_c$), with an exponent $\cong -0.5$ (average of 10~realizations).
    (b)~$\langle \tau \rangle$ scales with system size~$N$, at criticality,
    with an exponent $\cong 1/3$ (average of 20 realizations up to $N = 90000$,
    and of 15 runs for $N = 122500$). These results are similar to those
    found for interdependent networks~\cite{zho-14}. Error bars are included
    for each point of both curves.} \label{fig5}
\end{figure}

Continuing the comparison between spatial and nonspatial networks, we
now observe how the CFO evolves while reaching the final state, close to
criticality. In Fig.~\ref{fig4} (a), we show the time evolution of $S$
for $\zeta = 100$ and several values of $p$, with $p \leq p_c$.
The total time of the cascade, $\tau$, increases as the system gets
closer to criticality (see also Fig.~\ref{fig5} (a)). These cascades also
show a plateau in $S$, where a microscopic amount of failures
(Fig.~\ref{fig4} (b)) keeps the cascades going on with a branching factor
$\eta \approx 1$ (Fig.~\ref{fig4}~(c)), for a number of time steps of
the order of $N^{1/3}$ (see Fig.~\ref{fig5} (b)). Due to finite-size
effects, this phase does not last forever and, eventually, the amount
of failed nodes starts to increase because of accumulated damage in the
system, leading to an abrupt collapse~\cite{li-12,zho-14}. In 
Fig.~\ref{fig4} (d) we show the spatiotemporal distribution of the failures
just above criticality, where failures spread at all times over the whole
network. This occurs because optimal paths that disappear after some
failures are likely to be replaced by paths that pass through distant nodes
due to long-range connections, and then these distant nodes become overloaded.

The process for spatial networks ($\zeta =~3$, Fig.~\ref{fig4} (e)-(h))
is strikingly different. Since the typical length of links is short (compared
to $L$), initial failures due to overloads may concentrate and spread 
radially to close neighbors (Fig.~\ref{fig4} (h)). Eventually, near
criticality, overloads and failures create a hole of failed nodes within the functional giant component, which grows spontaneously and spreads throughout the entire system, causing its collapse. This phenomenon is known as nucleation, and it has also been observed in interdependent lattices with 
finite-length dependency links \cite{li-12,dan-14} and in spatial multiplex 
networks~\cite{dan-16,vak-17}. In addition, the complete disintegration of 
the giant component develops in a prolonged time interval with a relatively 
short plateau stage and a more gradual collapse (in contrast to weakly embedded systems, as seen in Fig.~\ref{fig4} (a)).

\section{Discussion}

In this paper we study the critical behavior and exponents that characterize
the steady state and the dynamics of cascading failures due to overloads, 
governed by the ML model and triggered by randomly removing a fraction of links,
in both strongly and weakly spatially-embedded 2D networks, which have a typical
link length $\zeta$. 

For the weakly or nonembedded systems we observe a usual mixed-order
transition similar to that of interdependent random networks, with a
critical exponent value of $\beta~=~0.5$. Furthermore, fluctuations
of the quantities $p_c$ and $M(p_c)$ present exponents also in
agreement with those of interdependent networks. These exponents
characterize the correlation length and the fractal fluctuations of
the order parameter. In contrast, strongly embedded networks
do not show a curvature (singularity) in the order parameter near $p_c$,
but rather a linear decrease, as in interdependent spatial networks,
which is a characteristic of pure first-order transitions. Regarding
dynamical aspects near the transition, weakly and strongly embedded systems 
also show a strikingly different behavior. Studying the spatiotemporal
propagation of failures, we find that for large $\zeta$ the failures spread
through the whole network at all times. In contrast, for small $\zeta$,
initial failures are likely to initiate in a random location and propagate
to nearby sites, yielding to a nucleation spreading process that is
observed as well in spatial interdependent and multiplex networks~\cite{li-12}
(see also the recent study by Choi~{\it et al.}~\cite{cho-23}).

Our results regarding the temporal evolution of CFO and those corresponding to
critical exponents at the steady state, for both small and large $\zeta$, show a
remarkable similarity to those of pure percolation (in the absence of overloads) of 
interdependent networks, for short- and long-range dependencies, respectively.
Therefore, we suggest that the overload mechanism of failure propagation plays a
similar role to that of dependencies in networks, and that both systems may belong to
the same universality class.

We recognize that our study is limited to the model of cascading failures proposed by
Motter and Lai, in which the shortest or optimal paths play a crucial role in determining 
the loads of the nodes and thus intervene in the dynamics of the process. Further
research exploring diverse failure propagation models would help to extend the scope
of the results found in this paper. For instance, one interesting model to analyze
would be the direct current approximation of power grids, where nodes (generators,
loads, or transmission nodes) satisfy the Kirchhoff equation, which will be our aim
in a future study.

In addition, the time-costly numerical simulations that allowed us to carry out this
study also limit us in the amount of results that we can produce in a certain amount
of work time. In this way our results, which are only performed for spatial dimension
$d = 2$, represent an indication that overloads in networks and interdependent
networks belong to the same universality class. However, it would be interesting to
explore higher dimensions and analyze the dependence of the critical exponents on
$d$, and test whether higher dimensions show also a similar behavior to that of Ref.
~\cite{gros-22-b} (Figs. 2 (a) and 3 (a)) for the fluctuations of the critical quantities $p_c$
and $M_c$. Related to this, and considering our limitation to study systems up to a
certain size, an analysis on how critical exponents approach the values obtained in
this work as the size of the system increases would make our results more robust.

Cascading failures due to overloads can dramatically alter the functioning of 
relevant infrastructures (e.g., electrical power grids, and communication and transportation
networks). Thus, researchers from various disciplines are interested in developing
a well-founded framework for understanding how such catastrophic processes behave
and what are their microscopic origin. We find that when long-range links appear the
cascade of failures occurs throughout the system, while for short-range links (with
respect to the system size) a nucleated damage occurs and propagates radially
throughout the system. Our study could therefore be useful for devising and building
more resilient infrastructures, in order to avoid or mitigate such catastrophic
breakdowns.

\section{Acknowledgements}

I. A. P., C. E. L. and L. A. B. wish to thank to UNMdP (EXA 1056/22), FONCyT
(PICT 1422/2019) and CONICET, Argentina, for financial support. S. H. wishes
to thank the Israel Science Foundation (Grant No. 189/19), the Binational 
Israel-China Science Foundation (Grant No. 3132/19), the NSF-BSF (Grant No. 
2019740), the EU H2020 project RISE (Project No. 821115), the EU H2020 DIT4TRAM,
and the EU H2020 project OMINO (Grant No. 101086321) for financial support. 
This research was supported by a grant from the United States-Israel Binational
Science Foundation (BSF), Jerusalem, Israel (Grant No. 2020255).


%

\clearpage
\section{Supplementary Information}
\setcounter{figure}{0}

\begin{figure}[hbt]
    \subfloat{\begin{overpic}[width = 0.4\textwidth, height = 0.3\textwidth]{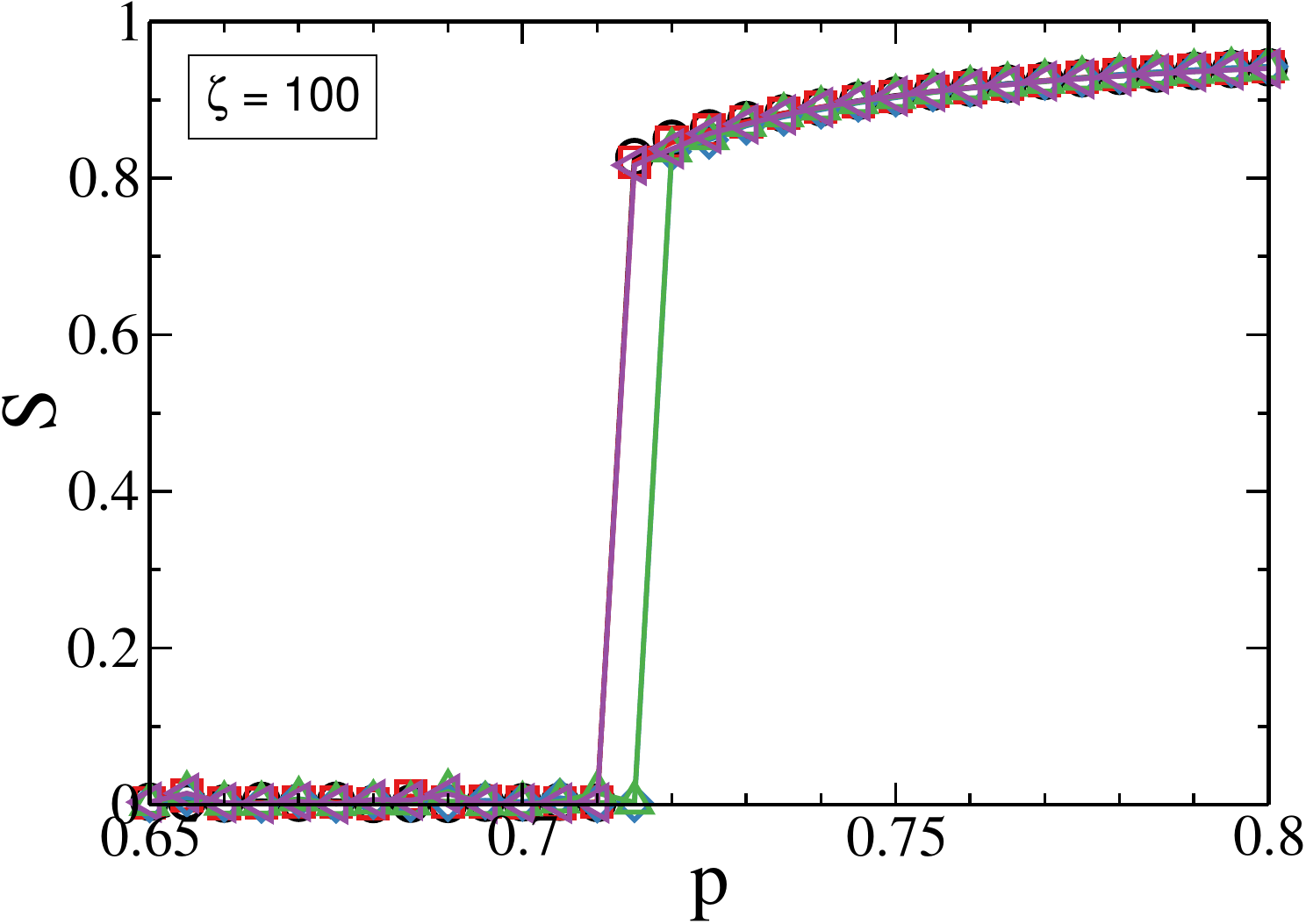}
        \put(0,0){\bf{(a)}}
    \end{overpic}} \\
    \subfloat{\begin{overpic}[width = 0.4\textwidth, height = 0.3\textwidth]{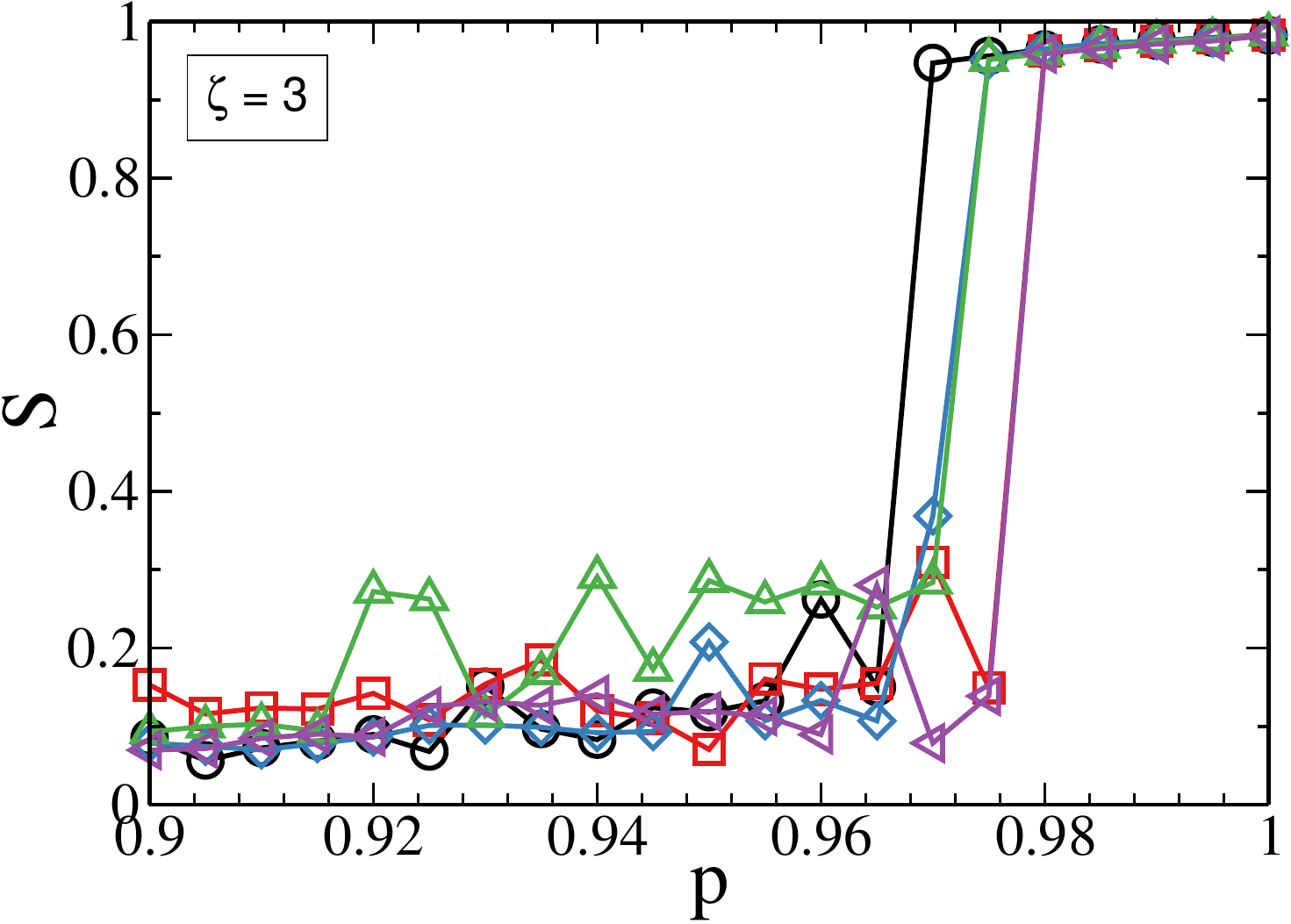}
        \put(0,0){\bf{(b)}}
    \end{overpic}}
    \caption{Individual realizations of the final size of the functional giant component,
    $S$, as function of the fraction $1 - p$ of randomly removed links. Results correspond
    to networks with (a)~$\zeta = 100$ and (b)~$\zeta = 3$, and they depict the different
    abrupt nature of the transitions. The remaining parameters are the same as in
    Fig.~\ref{fig2}.}
    \label{fig6}
\end{figure}

\end{document}